\documentclass[a4paper,fleqn]{cas-sc}

\usepackage[numbers]{natbib}

\def\tsc#1{\csdef{#1}{\textsc{\lowercase{#1}}\xspace}}
\tsc{WGM}
\tsc{QE}
\tsc{EP}
\tsc{PMS}
\tsc{BEC}
\tsc{DE}

\begin{document}
\let\WriteBookmarks\relax
\def\floatpagepagefraction{1}
\def\textpagefraction{.001}

\shorttitle{Kutta condition for potential flow}

\shortauthors{C Lozano and J Ponsin}

\title [mode = title]{A variational formulation of the adjoint Kutta condition in potential flow}
\tnotemark[1]

\tnotetext[1]{The research described in this paper has been supported by INTA under grant IDATEC (IGB21001).}
%
%

%
\author[1]{Carlos Lozano}
\cormark[1]

\fnmark[1]

\ead{lozanorc@inta.es}


\credit{Conceptualization (equal); Formal analysis (equal);
Investigation (equal); Writing (equal)}

\affiliation[1]{organization={National Institute of Aerospace Technology (INTA)},
    country={Spain}}

\author[1]{Jorge Ponsin}
\fnmark[1]

\credit{Conceptualization (equal); Formal
analysis (equal); Investigation (equal); Writing (equal)}

\cortext[cor1]{Corresponding author}

\fntext[fn1]{Theoretical and Computational Aerodynamics Group.}


\begin{abstract}
We give a variational formulation of the continuous adjoint Kutta condition for two-dimensional subcritical potential flow, with emphasis on the Kutta condition and the role of the wake. We show that the adjoint Kutta condition can be imposed by a penalty term evaluated at the trailing edge, with the corresponding Lagrange multiplier determined by stationarity of the Lagrangian with respect to circulation, and that a wake treatment is not required. Some of the implications of these results for adjoint consistency are also briefly discussed.
\end{abstract}




\maketitle

\section{Introduction}

The adjoint potential method remains an important tool in aerodynamic shape design and error analysis \cite{galbraith2017,crovato2023}. 
However, a significant gap remains in the characterization of the continuous adjoint PDE and boundary conditions for the linearized
incompressible or full potential equations. The difficulties are related to the formulation of the Kutta
condition and wake boundary conditions.

Here we address the issue in two dimensions. In potential flow theory, the circulation appears as a global degree of freedom that is not directly constrained by the governing partial differential equation, requiring the introduction of an additional constraint in both the flow and adjoint formulations. For flows around shapes with sharp trailing edges, the Kutta condition selects the physically admissible solution by fixing the global circulation to a particular value that removes the trailing-edge singularity. In addition, in two-dimensional formulations based on the velocity potential, a nonzero circulation renders the potential multi-valued, and a branch cut (wake cut) is introduced to restore single-valuedness. While this Kutta-wake construction is standard in the primal problem and is directly inherited by discrete adjoint implementations via algorithmic differentiation, it complicates the formulation of the continuous adjoint problem. 

This difficulty is not merely technical. A complete continuous sensitivity formulation requires a consistent adjoint treatment of the Kutta condition. Existing continuous adjoint approaches address this issue in different ways. Early formulations remove the issue at the outset by fixing the circulation a priori  \cite{Jameson1988}. More general treatments  allow circulation to vary, but they require adjoint constraints along the wake cut, resulting in an implicit adjoint wall boundary condition \cite{kuruvila1995}. Recently, Anevlavi et al. \cite{anevlavi2021} used a source-vorticity primal solver---which avoids a wake cut entirely---and imposed the adjoint Kutta condition as a local penalty term with the same multiplier used to enforce the field equations. That multiplier was then set to zero at the trailing edge to cancel the resulting variational boundary terms, thereby removing the sensitivity to circulation from the adjoint field. On the other hand, analytic adjoint solutions obtained in \cite{Lozano2025a,Lozano2025b} show that the Kutta condition gives rise to a singular contribution localized at the trailing edge. It was shown in \cite{Lozano2025a,Lozano2025b} that these singular contributions can be generated if certain terms are added to the Lagrangian; in that setting, those terms were interpreted as singular contributions to the linearized cost function rather than as variational constraints. The clarification of this point is one of the main motivations of the present paper. 

The purpose of the present note is to provide a complete variational formulation. We show that the adjoint Kutta condition can be imposed by a penalty term evaluated at the trailing edge, and that the associated Lagrange multiplier is fixed by stationarity of the Lagrangian with respect to circulation. The resulting boundary condition contains a localized singular term that reproduces, within a variational framework, the structure identified in analytic adjoint solutions. In addition to yielding a compact continuous formulation, this viewpoint also clarifies how dual consistency should be interpreted in this setting.

\section{Potential flows}
The adjoint equations can be used in many different contexts. For the sake of definiteness, we will restrict ourselves to the problem of computing the flow sensitivities of the aerodynamic force (projected along a certain direction $\vec{d}$) exerted on a two-dimensional profile $S$. We assume that the flow is inviscid and irrotational. Such a flow and, correspondingly, its adjoint problem, can be formulated using a velocity potential or a stream function. However, we limit our analysis to the velocity potential for two  reasons. First, in compressible flows, the relationship between the stream function and the density is implicit, making its application computationally cumbersome, which is why it is generally avoided in numerical solvers. Second, and foremost, only the potential formulation suffers from the wake cut issue, making its adjoint boundary treatment uniquely problematic.

We first consider the incompressible case, where exact solutions can be obtained using complex variable techniques, and after that we will address the generalization to compressible subcritical flows.

\subsection{Incompressible flows}
For incompressible flows, the domain $\Omega$ exterior to the profile $S$, with complex coordinate $z$, is conformally mapped onto the domain exterior to a circle of radius $R$, with complex coordinate $\zeta$. If $\Phi(\zeta)$ is the complex potential in the circle plane, then $\Phi(\zeta(z))$ is the corresponding potential in the airfoil plane and the complex velocities are related as $w_{airfoil} = w_{circle}/h$. The conformal mapping has a critical point at sharp trailing edges $z=z_{te}$ where $h=dz/d\zeta = 0$. Examining the expression for the complex velocity, the Kutta condition requires that $w_{circle}=0$ at $\zeta(z_{te})$ to avoid an unphysical singularity. In terms of the potential, this condition is $\partial_t\phi_{circle}=0$, which maps to $h \partial_t\phi$ on the airfoil, where $\partial_t$ is the tangent derivative along the profile or the circle. The Kutta condition can thus be imposed as $h \partial_t\phi = 0$ in the airfoil plane. Notice that the wake cut has been excluded from the analysis. As shown below, the above condition is sufficient to correctly reproduce the analytic adjoint solution.  

As stated above, the objective function is the aerodynamic force coefficient in a direction $\vec{d}$,
\begin{equation}
I=\frac{1}{c_{\infty}}\int_{S}(p-p_{\infty})(\hat{n}\cdot\vec{d})ds=-\frac{1}{2 c_{\infty}} \int_{S} \rho_{\infty}(\nabla\phi)^2(\hat{n}\cdot\vec{d})ds
\label{eq666}
\end{equation}
where $\rho_{\infty}$ is the (constant) fluid's density, $p$ is the pressure and $c_{\infty}$ is a normalization coefficient rendering $I$ dimensionless. Equation (\ref{eq666}) measures drag when $\vec{d}$ is parallel to the free-stream direction and lift when it is orthogonal thereto. We wish to compute the sensitivity of (\ref{eq666}) to changes in the flow. Using adjoint variables $\tilde\phi$ and $\tilde{\phi}_{te}$ to enforce the flow equations and the Kutta condition at the trailing edge, respectively, the augmented objective function (Lagrangian) is  
\begin{equation}
\begin{aligned}
L=&-\frac{1}{2 c_{\infty}} \int_{S} \rho_{\infty}(\nabla\phi)^2(\hat{n}\cdot\vec{d})ds -\int_{\Omega}\rho_{\infty}\tilde{\phi}\nabla^2\phi d\Omega  +\tilde{\phi}_{te}h\partial_{t}\phi|_{te} = \\&-\frac{1}{2 c_{\infty}} \int_{S} \rho_{\infty}(\nabla\phi)^2(\hat{n}\cdot\vec{d})ds -\int_{\Omega}\rho_{\infty}\tilde{\phi}\nabla^2\phi d\Omega  +\int_{S}\delta(s)\tilde{\phi}_{te} h\partial_{s}\phi ds
\label{eq667}
\end{aligned}
\end{equation}
In (\ref{eq667}), $\delta(s)$ is a Dirac delta function and we assume that the trailing edge is located at $s = 0$ without loss of generality. Linearizing (\ref{eq667}) with respect to perturbations in the flow solution yields, after integration by parts and rearrangement, 
\begin{equation}
\begin{aligned}
\delta L=&\frac{1}{c_{\infty}}\int_{S}\rho_{\infty}\partial_{s}(\partial_{s}\phi(\hat{n}\cdot\vec{d}))\delta\phi ds-\int_{\Omega}\rho_{\infty}\nabla^{2}\tilde{\phi}\delta\phi d\Omega-\int_{S}\rho_{\infty}\tilde{\phi}\partial_{n}\delta\phi ds+\int_{S}\rho_{\infty}\partial_{n}\tilde{\phi}\delta\phi ds\\
&-\int_{S_{\infty}}\rho_{\infty}\tilde{\phi}\partial_{n}\delta\phi ds+\int_{S_{\infty}}\rho_{\infty}\partial_{n}\tilde{\phi}\delta\phi ds-\int_{S}\partial_{s}(\delta(s)\tilde{\phi}_{te} h)\delta\phi ds
\label{eq669}
\end{aligned}
\end{equation}

The adjoint equation and boundary conditions follow directly from (\ref{eq669}). The adjoint potential $\tilde{\phi}$ obeys Laplace's equation $\nabla^{2}\tilde{\phi} = 0$ in $\Omega$, vanishes in the farfield and obeys the following wall boundary condition

\begin{equation}
\partial_{n}\tilde{\phi}|_{S}=-\partial_{s}(\partial_{s}\phi(\hat{n}\cdot\vec{d}))/{c_{\infty}}+\rho_{\infty}^{-1}\partial_{s}(\tilde{\phi}_{te} h\delta(s))
\label{eq671}
\end{equation}
where the derivative of the Dirac delta function must be understood in the distributional sense, and its discrete counterpart arises through consistent integration over control volumes or elements. 

What remains is to determine the value of the adjoint multiplier  $\tilde{\phi}_{te}$ associated with the Kutta constraint. This is achieved by noting that the Lagrangian must be stationary with respect to perturbations in the global circulation $\Gamma$, with all other quantities fixed (geometry, angle of attack, density, free-stream velocity):
\begin{equation}
\frac{\delta L}{\delta\Gamma}=0
    \label{Stationary}
\end{equation}
This condition follows from the fact that, in the augmented variational formulation, the circulation remains an independent degree of freedom. Although the Kutta condition ultimately determines its value, this constraint is enforced through a Lagrange multiplier rather than being imposed a priori. Consequently, variations with respect to the circulation are admissible, and stationarity of the Lagrangian requires (\ref{Stationary}).  

Let $\mathcal{K} = h\partial_{t}\phi|_{te}$ denote the Kutta penalty term. Since a pure circulatory perturbation satisfies Laplace's equation, the variation of the domain integral in (\ref{eq667}) vanishes, leaving:
\begin{equation}
    \frac{\partial L}{\partial \Gamma}=\frac{\partial I}{\partial \Gamma} + \tilde{\phi}_{te} \frac{\partial \mathcal{K}}{\partial \Gamma} = 0
    \label{eq:zero_circulation}
\end{equation}
The consequence of (\ref{eq:zero_circulation}) depends on whether the objective function is drag or lift. Potential flows have zero drag. Hence, $\partial I/\partial \Gamma=0$ in that case, so the multiplier for the drag-based adjoint equation $\tilde{\phi}_{te}=0$ and the singular term drops out of (\ref{eq671}). On the other hand, for lift we have $I=C_L = -\rho_{\infty} q_\infty \Gamma/ c_{\infty}$ and, thus:
\begin{equation}
    \frac{\partial I}{\partial \Gamma} = -\rho_{\infty} q_\infty/ c_{\infty}
\end{equation}
In the mapped circle plane of radius $R$, the tangential velocity induced by circulation at the rear stagnation point ($\theta = 0$) is $v_\theta = \Gamma / (2\pi R)$, yielding:
\begin{equation}
    \frac{\partial \mathcal{K}}{\partial \Gamma} = \frac{1}{2\pi R}
\end{equation}
Substituting these sensitivities into the stationarity condition provides the exact value for the multiplier:
\begin{equation}
    -\rho_{\infty} q_\infty/ c_{\infty} + \tilde{\phi}_{te} \left( \frac{1}{2\pi R} \right) = 0 \implies \tilde{\phi}_{te}=2\pi R \rho_{\infty} q_\infty/ c_{\infty}
    \label{eq673}
\end{equation}

The solution to Laplace's equation for the lift-based adjoint variable with boundary condition (\ref{eq671}) and condition (\ref{eq673}) and vanishing in the farfield is \cite{Lozano2025a}
\begin{equation}
\tilde{\phi}(z) = v(z) \cos \alpha - u(z) \sin \alpha + q_{\infty} \Upsilon^{(1)}(z(\zeta)) / c_{\infty}
\label{eq1}
\end{equation}
where $(u, v)$ are the Cartesian components of the velocity, $q_{\infty}(\cos \alpha, \sin \alpha)$ is the free-stream velocity and $\Upsilon^{(1)} \sim \text{Im}(1/(\zeta-R))$ is the conjugate Poisson kernel for the Laplacian on the exterior of the circle. (The stream function formulation is the harmonic conjugate of the above, so the corresponding Kutta condition leads to a solution that contains the Poisson kernel). From a physical viewpoint, the Kutta function $\Upsilon^{(1)}$ arises as the amount of circulation required to maintain the Kutta condition at the trailing edge or rear stagnation point of the profile. $\Upsilon^{(1)}$ is singular at the rear stagnation point of the circle, giving rise to the adjoint singularity at the trailing edge of the profile described by (\ref{eq671}).

\subsection{Compressible flows}
The compressible problem does not admit a closed-form solution. We can nevertheless proceed along the same lines as in the previous case, except that, instead of a conformal mapping, we use Bers’ quasi-conformal transformation \cite{bers1954}, which maps the subsonic potential flow around the profile $S$ to an incompressible flow around a unit circle. Most of the technical ingredients needed in the compressible case, together with some indirect numerical support, are developed in \cite{Lozano2025b}. Here we extract only the parts needed for the present variational formulation.

As in the incompressible case, the Kutta condition is incorporated into the adjoint formulation by an adjoint weighted penalty term 
\begin{equation}
L =\int_{S}{c_{\infty}}^{-1}p(\hat{n}\cdot\vec{d})ds-\int_{\Omega}\tilde{\phi}\nabla\cdot(\rho\nabla\phi)d\Omega+\tilde{\phi}_{te}h\partial_{t}\phi|_{te}
\label{eq71}
\end{equation}
where  $h$ is the modulus of the quasi-conformal mapping. Linearizing eq. (\ref{eq71}) with respect to flow variations, integrating by parts and rearranging yields the equation and boundary conditions obeyed by the adjoint variable
\begin{equation}
\partial_{j} \left( \rho \left( \delta_{ij} - \frac{1}{a^{2}} u_{i} u_{j} \right) \partial_{i} \tilde{\phi} \right) = 0
\label{eq6}
\end{equation}

\begin{equation}
\partial_{n}\tilde{\phi}_{S}=-\rho^{-1}\partial_{s}(\rho(\hat{n}\cdot\vec{d})\partial_{s}\phi)/c_{\infty}+\rho^{-1}\partial_{s}(\tilde{\phi}_{te}h\delta(s))
\label{eq66}
\end{equation}
with $\tilde{\phi}$ vanishing in the farfield. In (\ref{eq6}), $a$ is the local speed of sound, $i,j=1,2$ denote Cartesian components and summation over repeated indices is understood. As before, the derivative of the Dirac delta function must be understood in the appropriate sense.

Bers' mapping depends on the flow, so linearization of eq. (\ref{eq71}) produces an additional term $\tilde{\phi}_{te}\delta h\partial_{t}\phi|_{te}$. Because of the Kutta condition, this term vanishes regardless of $\delta h$ except for cusped trailing edges. In that case, the original and perturbed geometries are both cusped or nearly cusped, so their corresponding moduli both vanish at the trailing edge with exponents equal or close to $1/2$ \cite{bers1954}, so $\delta h_{te} \to 0$.

The value of $\tilde{\phi}_{te}$ can again be fixed by enforcing (\ref{Stationary}). Since $C_L = -\rho_\infty q_\infty \Gamma/ c_{\infty}$ also 
 holds for compressible flows \cite{finn1958}, the circulation mode (the linearized perturbation associated to a circulation perturbation) obeys the linearized potential equation, and $h\partial_t\phi^{airfoil}=\frac{1+\tilde\rho}{2\tilde\rho}\partial_t\phi^{circle}$, it can be shown that (\ref{Stationary}) yields $\tilde{\phi}_{te} = 0$ for the drag-based adjoint and  
\begin{equation}
\tilde{\phi}_{te}=\frac{4\pi \rho_\infty q_\infty}{c_\infty} \frac{\tilde{\rho}_{te}}{1+\tilde{\rho}_{te}} 
\label{eq75}
\end{equation}
for the lift adjoint, where $\tilde\rho=\rho/\rho_{\infty}$. Solving (\ref{eq6}) with conditions (\ref{eq66}) and (\ref{eq75}) is a formidable task. Inspecting the boundary condition, and by analogy with the incompressible case, one would expect the solution to possess a regular part and a singular part stemming from the Dirac delta forcing in the boundary condition. Using the Green's function approach \cite{giles2001}, it is possible to obtain a closed-form expression for the regular part, with the singular part being given in terms of a function $\tilde{\Upsilon}^{(1)}$ resulting from solving the generalized Laplace equation (\ref{eq6}) with the singular forcing but whose explicit form is not known in general. For lift, the result is \cite{Lozano2025b}   
\begin{equation}
\tilde{\phi}_{L}=\frac{1}{c_{\infty}}(v\cos\alpha-u\sin\alpha)+\frac{q_{\infty}}{c_{\infty}}\tilde{\Upsilon}^{(1)} 
\label{analyticAdjSoln}
\end{equation}

The solution is formally identical to the incompressible case (\ref{eq1}), but now $\tilde{\Upsilon}^{(1)}$ is the generalized conjugate Poisson kernel for the generalized Laplacian (\ref{eq6}). $\tilde{\Upsilon}^{(1)}$ is related to the boundary tangent derivative of the Neumann Green's function $\hat{G}_{N}$ of (\ref{eq6})
\begin{equation}
\tilde{\Upsilon}^{(1)}(\vec{x})=4\pi\rho_{\infty}\frac{\tilde{\rho}_{te}}{1+\tilde{\rho}_{te}}h_{te}\partial_{s}\hat{G}_{N}(\vec{x},\vec{y}(s))_{s=s_{te}}
\label{eq70}
\end{equation}
$\tilde{\Upsilon}^{(1)}$ can be shown to obey a Neumann condition $\partial_{n} \tilde{\Upsilon}^{(1)}|_{wall} = 0$ on the airfoil profile--excluding the trailing edge, where it obeys the singular part of (\ref{eq66})-- and reduces to $\Upsilon^{(1)}$ in the incompressible limit.

The stream function formulation is the A-harmonic conjugate of the above \cite{Astala2009}. Hence, the corresponding singular function can be written in terms of the generalized Poisson kernel (the boundary conormal derivative of the Dirichlet Green's function) for the conjugate generalized Laplacian governing linearized perturbations to the stream function.


\subsection{Dual consistency and localized adjoint forcing}

Dual or adjoint consistency is a property of discretizations of the primal equation such that the corresponding discrete adjoint equation is a consistent approximation of the corresponding continuous adjoint problem. Adjoint consistent discretizations are relevant because they have optimal grid convergence properties. Likewise, adjoint-based error estimation requires adjoint consistency. In either case, a complete continuous adjoint formulation is required to show adjoint consistency. In the present setting, this requires the identification of the correct continuous adjoint formulation of the Kutta condition.

Traditional continuous adjoint treatments of potential flow enforce the Kutta condition through non-local constraints applied along the wake cut. Analytic adjoint potential solutions, on the other hand, show that the continuous adjoint problem contains a singular contribution localized at the trailing edge that can be obtained without incorporating the wake cut into the variational formulation. 

Checking the dual consistency of a primal discretization amounts to showing that the discrete adjoint formulation reproduces the continuous adjoint boundary-value problem derived above, including the singular boundary behavior at the trailing edge induced by the Kutta condition. In potential flow discretizations containing wake cuts, circulation is represented through a potential jump across the wake, whose value is determined by the trailing-edge condition. Any discrete adjoint system obtained by algorithmic differentiation therefore inherits the discrete structures introduced by that wake representation and by the Kutta enforcement. Whether such structures are compatible with the continuous adjoint problem must be analyzed on a case-by-case basis. On the other hand, a formulation without wake cuts like the one proposed in \cite{anevlavi2021} appears, a priori, to be a natural candidate for an adjoint-consistent discretization, but this requires further investigation 


\section{Conclusion}

A variational formulation of the continuous adjoint Kutta condition for two-dimensional subcritical potential flow has been presented. The Kutta condition is enforced by a penalty term evaluated at the trailing edge, and the associated Lagrange multiplier is determined by enforcing stationarity of the Lagrangian with respect to the global circulation. In this way, one obtains a continuous adjoint formulation without introducing a wake cut into the variational problem. The main point is that the singular forcing identified in previous analytic work follows naturally from the variational construction.

These results also shed light on the issue of dual consistency. The relevant continuous problem is clearly identified: it consists of the adjoint equation and boundary conditions, including the singular trailing-edge forcing. A dual-consistent discretization should therefore lead to a discrete adjoint formulation that is consistent with that continuous problem. In discretizations containing wake cuts, this requires a careful assessment of whether the discrete structures introduced by the wake and Kutta conditions are consistent with the continuous adjoint problem.

\printcredits

\bibliographystyle{model1-num-names}


\bibliography{refs}


\end{document}